\documentclass[aps,prb,twocolumn]{revtex4}

\usepackage[dvips]{graphicx,color}

\input epsf.sty

\begin{document}


%
%

\title{Dielectric properties and lattice dynamics of 
Ca-doped K$_{0.95}$Li$_{0.05}$TaO$_{3}$}

\author{S. Wakimoto}
\email[Corresponding author: ]{wakimoto.shuichi@jaea.go.jp}
\affiliation{ Quantum Beam Science Directorate, 
   Japan Atomic Energy Agency, Tokai, Ibaraki 319-1195, Japan. }

\author{G. A. Samara, R. K. Grubbs and E. L. Venturini}
\affiliation{ Sandia National Laboratories, 
   Albuquerque, New Mexico 87185, USA }


\author{L. A. Boatner}
\affiliation{ Oak Ridge National Laboratory, 
   Oak Ridge, Tennessee 37831, USA }

\author{G. Xu and G. Shirane~\footnote{Deceased.}}
\affiliation{ Department of Physics, Brookhaven National Laboratory,
   Upton, New York 11973, USA }


\author{S.-H. Lee}
\affiliation{ NIST Center for Neutron Research, National Institute of
  Standards and Technology, Gaithersburg, MD 20899 }
\affiliation{ Department of Physics, University of Virginia, 
  Charlottesville, VA22904, USA }

\date{\today}

\begin{abstract}

Relaxor behavior and lattice dynamics have been studied by employing 
dielectric measurements and neutron scattering methods 
for a single crystal of K$_{1-x}$Li$_x$TaO$_3$ $(x=0.05)$, 
where a small amount of a Ca impurity ($\sim 15$~ppm) 
was incorporated during the single crystal growth procedure.  
The dielectric constant $\epsilon'(\omega, T)$ 
shows qualitatively similar behavior to that of Ca-free KLT 
with $x=0.043$ with both compositions exhibiting relaxational
properties with no evidence for a ferroelectric transition.
The absolute value of $\epsilon'(\omega, T=0)$ for the present 
crystal is larger by an order of magnitude than that of the 
Ca-free sample due to charge carriers 
induced by the Ca doping.  
This large value is shown to be due to a Maxwell-Wagner 
relaxation process associated with the low temperature ($< 8$~K) 
activation of frozen electronic carriers.
The dielectric loss tangent 
$\tan \delta$ reveals three Debye-like relaxations with 
Arrhenius activation energies of $80$, $135$, and $240$~meV 
that are assigned to Li$^+$ dipoles, Ca$^{2+}$-related relaxation 
and the Li$^+$-Li$^+$ dipolar pairs, respectively.
In the neutron scattering results, diffuse scattering ridges appear around the 
nuclear Bragg peaks along the [100] direction below $\sim 150$~K 
and phonon line broadening features start to appear at even higher 
temperatures suggesting that polar nano-regions 
(PNR's) start to form at these temperatures.
These results are supported by the dielectric data that reveal 
relaxor behavior starting at $\sim 200$~K on cooling.
From analyses of the diffuse intensities at different zones, 
we have derived atomic displacements in the PNR's.  The results 
suggest that the displacements include a uniform 
phase shift of all of the atoms in addition to the atomic 
displacements corresponding to a polarization vector of the 
transverse optic soft ferroelectric mode, a finding that is 
analogous to that in the prototypical relaxor material 
Pb(Mg$_{1/3}$Nb$_{2/3}$)O$_3$.  

\end{abstract}


\maketitle

\section{Introduction}

Relaxor ferroelectrics that have a relatively large dielectric 
permittivity over a wide temperature range have attracted a significant amount of 
attention due to their high potential for device applications 
as well as their scientific challenges.  It is now widely believed 
that relaxor behavior, whose universal signature is a frequency($\omega$)-dependent 
peak in the dielectric constant $\epsilon'(\omega, T)$, arises from the relaxational 
nature of randomly-oriented polarized nano-regions (PNR's).  
However, the mechanism for the PNR formation is not fully understood 
yet.  In particular, unlike the case for displacive ferroelectric materials, 
possible contribution of the zone-center transverse optic (TO) soft 
phonons to the PNR condensation still remains to be clarified.

The relaxor material system K$_{1-x}$Li$_x$TaO$_3$ (KLT($x$)) provides a unique 
opportunity for systematic research ranging from a relaxor state in the material at low Li
concentrations to a ferroelectric state at high concentrations.  
The parent compound KTaO$_3$ is an incipient ferroelectric 
material that does not undergo a ferroelectric 
transition~\cite{Wemple_65} although the TO soft 
mode softens remarkably at low temperature.~\cite{Axe_70}  On substituting at the K$^{+}$ 
site, the much smaller Li$^{+}$ ion occupies an off-center 
positions resulting in a large dipole moment that is the precursor for
PNR formation on cooling.  At high Li concentration, the PNR's percolate, 
and the system achieves ferroelectric order.

Although the system KLT($x$) has been studied extensively by 
Raman scattering,~\cite{Raman,Vogt_new,Prater_81} and its soft mode 
character is well-established, only a few neutron scattering studies have been 
carried out mainly to explore the role of the soft mode of the host crystal 
in the formation and growth of the PNR's.
Yong {\it et al.}~\cite{Toulouse_00} have studied the 
diffuse neutron scattering in KLT that originates from 
PNR formation.  
They reported a strong diffuse intensity at (110) and its absence 
at both (100) and (200).
If we assume a PNR condensation from the TO soft mode as in 
the case of the ordinary ferroelectrics, then the structure factors 
for the diffuse and the inelastic scattering by the TO mode should agree.
However, it is found that the inelastic structure factor for the TO soft 
mode of KLT at (200) is much larger than that at (110) in 
disagreement with the diffuse scattering intensities.

Such disagreements 
between the diffuse intensities and inelastic structure factors 
have also been reported in the relaxor material
Pb(Mg$_{1/3}$Nb$_{2/3}$)O$_3$ (PMN).~\cite{Naberezhnov_99}
Moreover, in this latter case, the atomic displacements determined 
from the diffuse intensities at several zones do not conserve the 
center-of-mass (CM).~\cite{Vakhrushev_95}
These facts suggest an apparent disconnection between the PNR 
condensation and the soft phonon mode in PMN.  However, 
Hirota {\it et al.}~\cite{Hirota_02} have reconciled these 
two observations by introducing a model of the soft mode condensation 
that incorporates an {\it additional uniform phase shift}, in which the total 
atomic displacements can be divided into two components.  
One component originates from the normal condensation of the TO 
soft mode that conserves the CM, and the other component represents a 
uniform shift of all of the atoms in the polarization direction.

In the present paper, we report and discuss dielectric 
and neutron scattering experiments performed on KLT with 
$x=0.05$.
The results provide detailed views of the response and physics of 
this crystal.  An additional objective is to determine if the 
anomalous lattice dynamical features observed in the prototypical 
relaxor PMN are common to other ABO$_3$ relaxors.  This comparison 
offers us a unique opportunity to compare two different types of 
relaxors, one is KLT in which 
off-center impurity ions trigger a local polarization (PNR's) on the 
incipient ferroelectric background, and the other is PMN in which 
random occupation of the $B$-sites by two different ions with different 
valences produces the disorder that breaks long-range ferroelectric (FE) 
correlations and order resulting in the formation of a relaxor (R) state.  


\section{Experimental details}

The KLT samples used in the present study were $<100>$ oriented 
single crystals cut from one boule that was grown by 
solidification from the melt.  The boule was nominally undoped, 
but our dielectric measurements (to be presented below) suggested 
the presence of an unintentional dopant leading to some electronic 
conduction and a high dielectric loss.  Subsequent analysis 
revealed the presence of 
only
Ca impurity (at a level of $\sim 15$~ppm) 
in the starting 
powder used in the crystal growth.~\cite{Lynn_note}  Clearly, some of the Ca was 
incorporated into the crystal, but at a sufficiently low level 
that was difficult to quantify.  The crystal was colorless 
with no indication of any significant free carrier absorption 
that usually produces a blue coloration in more highly conducting crystals of this type.  
Although the Ca dopant had a definite influence on the dielectric 
properties, normally it would not be expected to have a significant influence 
on the lattice dynamics measured by neutron scattering.

It is well established that not all of the Li in the starting 
mixture of the oxides is incorporated into the growing crystal during 
the solidification process due to segregation coefficient effects.  
It is estimated that 35~\% of the Li content in the growth charge 
is incorporated.~\cite{Klink_84}
The Li composition of our KLT crystal ($\simeq 5$ at.\%) was estimated 
from the amount of Li in the melt and from the established 
relationship between the peak temperature in the $\epsilon'(T)$ 
response and the Li content.~\cite{Samara_SSP}  Henceforth we shall 
designate our composition as KLT(5):Ca.

For the dielectric measurements, the large (100) faces of the crystal 
were polished using colloidal silica and were subsequently 
covered with vapor-deposited chromium followed by gold.  
The sample was investigated by dielectric spectroscopy with 
measurements of the real ($\epsilon'$) and loss ($\tan \delta$) 
parts of the dielectric function performed as functions of 
temperature ($4 - 300$~K), frequency ($10^2 - 10^6$~Hz) and 
hydrostatic pressure ($0 - 7.5$~kbar).  The pressure system 
consisted of a compressor and an intensifier that fed the 
pressure-transmitting medium (helium gas) into a cell placed 
in a low temperature Dewar.  The pressure was measured using a 
calibrated manganin gauge to an accuracy of better than 4\%.

The neutron scattering experiments were performed using triple 
axis spectrometers SPINS at NIST, LTAS at JAEA, and TAS1 at 
JAEA.  The diffuse scattering was measured at SPINS and LTAS 
with an incident neutron energy of $E_i=5$~meV ($\lambda=4.04$~\AA) 
and a collimation sequence: Guide-80$'$-S-80$'$-open (S denotes 
the sample), and also at TAS1 with $E_i=14.7$~meV ($\lambda=2.36$~\AA) 
and 40$'$-40$'$-S-40$'$-80$'$.  The phonon modes were studied at 
TAS1 with a fixed final energy of $E_f=14.7$~meV and 
40$'$-40$'$-S-80$'$-open.  Beryllium and pyrolytic graphite filters 
were utilized for the measurements with $\lambda=4.04$~\AA~ and 
$\lambda=2.36$~\AA, respectively, to remove those neutrons with higher 
harmonic wavelength $(\lambda/2, \lambda/3, etc.).$

A KLT(5):Ca crystal with dimensions of $6 \times 10 \times 
13$~mm$^3$, cut from the same crystal boule that was used for 
the dielectric measurements, was prepared for the neutron 
scattering experiments.  The crystal was mounted in an unstrained manner on an 
aluminum post using aluminum foil, and the aluminum 
post was then masked by a cadmium plate to avoid any scattering 
from it.  The sample was sealed in an aluminum can filled 
with He gas, and placed in a He-gas closed-cycle refrigerator 
with the [100] and [010] axes in the horizontal scattering plane.
In the temperature range $10 \leq T \leq 295$~K, the crystal 
structure was cubic as evidenced 
in the neutron scattering resolution, with a lattice constant 
3.95~\AA, corresponding to a reciprocal lattice unit (r.l.u.) of 
1.59~\AA$^{-1}$.

\section{${\rm \bf Li}$ and ${\rm \bf Ca}$ substitution in ${\rm \bf KTaO}_3$}

Before presenting the dielectric and lattice dynamics 
properties, we briefly review in this section the effects of 
Li and Ca substitution on KTaO$_3$.

The substitution of Li$^{+}$ for K$^{+}$ in KLT raises an 
immediate issue.  The K$^{+}$ ion has a radius $r_i$ of 1.3~\AA~ and 
fits well in the lattice, occupying a centrosymmetric position 
in the oxygen ion cage that surrounds it in KTaO$_3$ (Fig. 1 (a)).  
The Li$^{+}$ ion, on the other hand, is too small ($r_i = 0.68$\AA), 
and the oxygen cage dimension is too large in comparison.  This is a circumstance that 
occurs relatively often in solids with the expected consequence 
that the small ion will shift to an off-center position.  
This is expected for Li in KLT, and indeed, this effect was observed early 
on.~\cite{Samara_SSP,Hochli_90,Yacoby_74}  The shift, as expected 
from the lattice symmetry, is in the [100] direction (Fig. 1 (b)) and 
is quite large (1.2~\AA).~\cite{Hochli_90}  
In this off-center position, the Li$^{+}$ ion has a relatively 
large dipole moment, which can interact with neighboring Li dipoles.  
Additionally, because of the high polarizability of the KTaO$_3$ host, 
the Li dipole can polarize adjacent unit cells resulting in polar 
nano-regions.  These are precisely the conditions that lead to 
dipolar glass and relaxor behavior.  Indeed, this is what is 
observed experimentally.~\cite{Samara_SSP,Hochli_90}

\begin{figure}
\centerline{\epsfxsize=3.3in\epsfbox{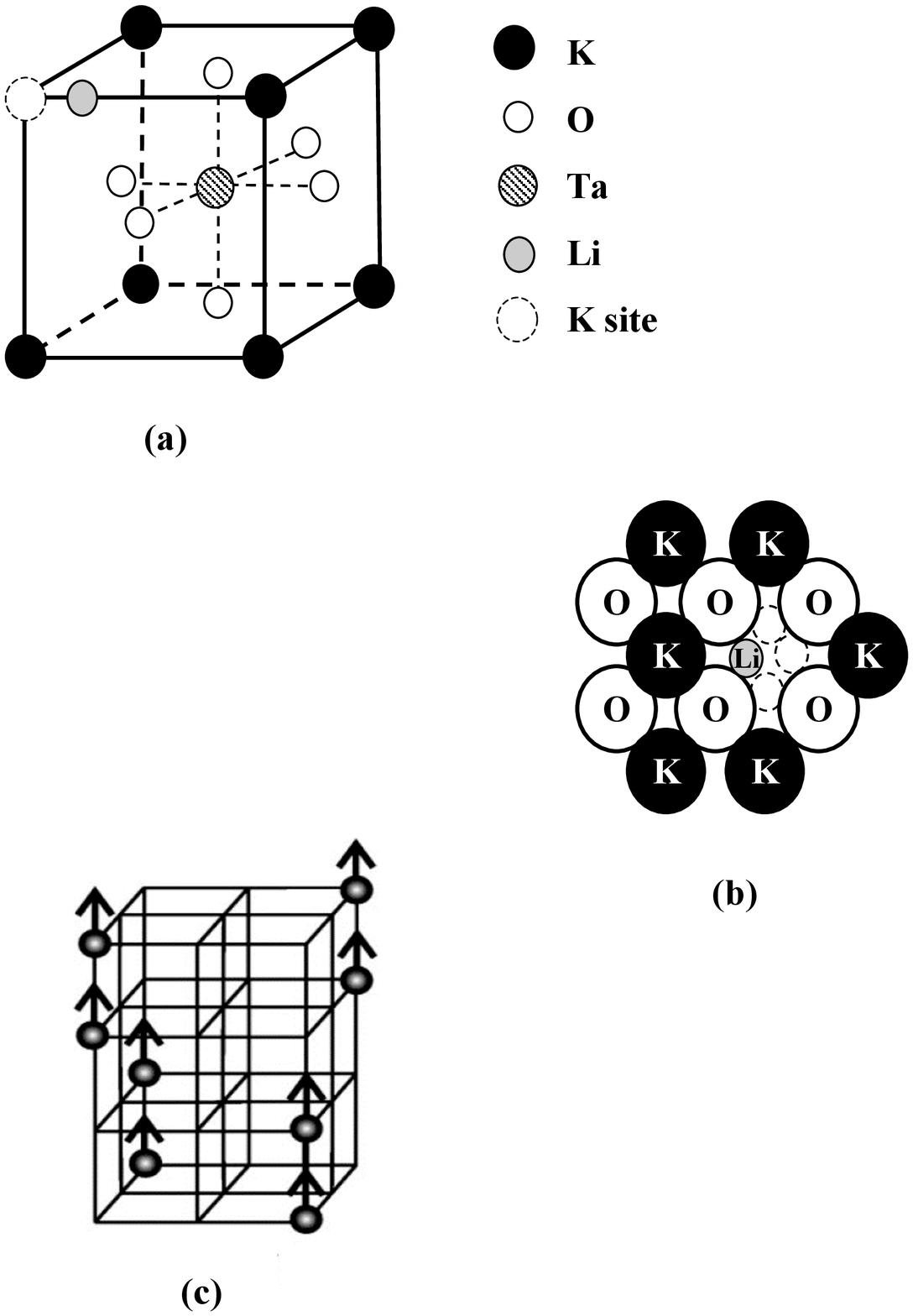}}
\caption{(a) The ABO$_3$ structure of KLT showing the off-center 
position of the Li ion at the K site.  The Li$^{+}$ hops among 6 
equivalent positions, 4 of which are shown in (b).  (c) shows the 
configuration of the co-linear Li$^{+}$-Li$^{+}$ ion pairs.}
\end{figure}

Leung~\cite{Leung_01} has performed a first-principles study of 
Ca$^{2+}$ substitution in KTaO$_3$.  Although his study is for 
KTaO$_3$ and not on KLT, we believe that the results are relevant 
to our Ca-doped KLT crystals.  The results show that Ca$^{2+}$ 
substitutes at both the K$^{+}$ and Ta$^{5+}$ sites.  
At the K$^{+}$ site, the Ca antisite, Ca$_{\rm K}^{+}$, would be 
expected to produce a K vacancy, but LeungÕs results find this 
circumstance highly unfavorable energetically.  Thus, the Ca$^{2+}$ 
simply donates an electron.  At the Ta$^{5+}$ site, the Ca$^{2+}$ 
binds to an oxygen vacancy (V$_{\rm O}$), thereby forcing the Ca$^{2+}$ 
off-center and resulting in a dipolar complex.  (Oxygen vacancies 
are formed on Ca$^{2+}$ substitution to preserve charge neutrality, 
and they are also native defects in perovskites).  The calculations 
predict that the activation energy for the reorientation of the 
(Ca-V$_{\rm O}$) dipole via oxygen-vacancy hopping within the first neighbor 
shell exceeds 2 eV.  This large energy precludes the direct 
involvement of the (Ca-V$_{\rm O}$) complex in the observed new relaxations 
that have much lower activation energies in the present KLT crystals. 
(See the next section.) 

Early work~\cite{Senhouse_66,Wemple_65} on Ca-doped KTaO$_3$ (KT) 
has suggested that the Ca is fully ionized at room temperature.  
Presuming that Ca$^{2+}$ substitutes at the K$^{+}$ site, Senhouse 
{\it et al.}~\cite{Senhouse_66} showed that the net ionized donor 
concentration is equal to the Ca concentration.  These results would 
suggest that most of the Ca$^{2+}$ resides at the K$^{+}$ site.  
However, impurities and defects (primarily oxygen vacancies that are 
prominent defects in ABO$_3$ oxides and in Ca-doped KT~\cite{Yacoby_74}) 
can also be sources of free electrons.  Wemple~\cite{Wemple_65} also 
showed that there is no ``freeze-out'' of the Ca-generated electrons in 
KT down to temperatures approaching liquid helium temperature.

When present in sufficient concentrations ($> \sim 10^{17} /cm^{3}$) 
the Ca-generated free electrons in Ca-doped KT and related crystals 
produce free carrier absorption and an associated blue coloration.  
In our present KLT(5):Ca crystal the Ca was present in small amounts, 
and consequently these crystals were colorless - containing relatively few free 
electrons and no noticeable free carrier absorption.  
However, this small amount of Ca had a significant influence on the 
amplitude of dielectric response and also a minor influence on the 
lattice dynamics.  (See later discussion.)

\section{Results and discussion}

\subsection{Dielectric response}

Figure 2 shows the dielectric response of our KLT(5):Ca crystal at 1~bar.  
The inset in Fig. 2(a) shows $\epsilon'(\omega,T)$ for an undoped 
4.3 at.\% Li KLT crystal~\cite{Prosandeev_01} - a composition whose 
dielectric response is seen to be very close to KLT(5).  The main 
overlapping peaks in this relaxational response are associated 
with the relaxations of isolated Li$^{+}$ dipoles (at the lower 
temperature) and with (Li$^{+}$ - Li$^{+}$) ion pairs with kinetic 
parameters $E \sim 80 (240)$~meV and $\omega_o \sim 1.5 \times 
10^{12}$~s$^{-1}$ ($7 \times 10^{14}$~s$^{-1}$), 
respectively.~\cite{Prosandeev_01,Doussineau_93,Toulouse_98JKPS}  
The relaxational dielectric response of our KLT(5):Ca crystal in 
Fig. 2(a) has some similarities to that in the inset, but there 
are important differences.

\begin{figure}
\centerline{\epsfxsize=3.3in\epsfbox{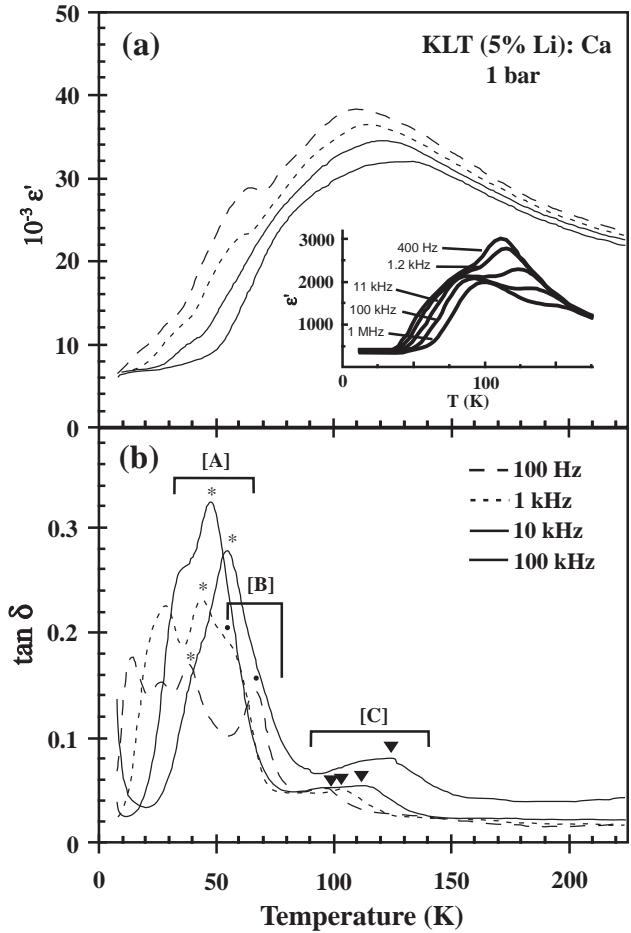}}
\caption{The 1~bar $\epsilon'$ and $\tan \delta$ responses of the 
KLT(5):Ca crystal.  The inset is the response of undoped KLT(4.3) 
according to Prosandeev {\it et al.} (Ref.~\onlinecite{Prosandeev_01}.)  
Three primary relaxations denoted by A, B and C are noted in (b).}
\end{figure}

Starting at low temperatures, we note that the amplitude of $\epsilon'$ 
below $\sim 20$~K ($\sim 6000$) is over an order of magnitude 
larger than that of the undoped KLT crystal of about the same Li 
concentration as shown in the inset in Fig. 2 (a).  This strong low-$T$ 
enhancement of $\epsilon'$ in KLT(5):Ca is due to the Ca$^{2+}$ 
doping.  A feature in the KLT(5):Ca data in Fig. 2, namely a 
relaxational feature below 10~K, points to the mechanism for 
this enhancement.  This feature is shown in Fig. 3.  The onset 
of this relaxation is reflected in the simultaneous highly 
dispersive decrease in $\epsilon'$ and increase in $\tan \delta$ 
at 1~bar.  This relaxation is accompanied by strong frequency 
dispersion in both $\epsilon'(T)$ and $\tan \delta(T)$, a step 
in $\epsilon'(T)$ with $\epsilon'$ increasing from $\sim 600$ 
at the lowest temperatures (the expected value for this composition in 
the absence of Ca [See inset in Fig. 2 (a)]) to $\sim 6000$ at 
the plateau region, and a large peak in $\tan \delta(T)$.  These 
features are the signature of a capacitive barrier layer that, 
for our present case, could be due to either a Schottky barrier 
or space charge accumulation at the metal contacts.~\cite{Grubbs_05}

\begin{figure}
\centerline{\epsfxsize=3.3in\epsfbox{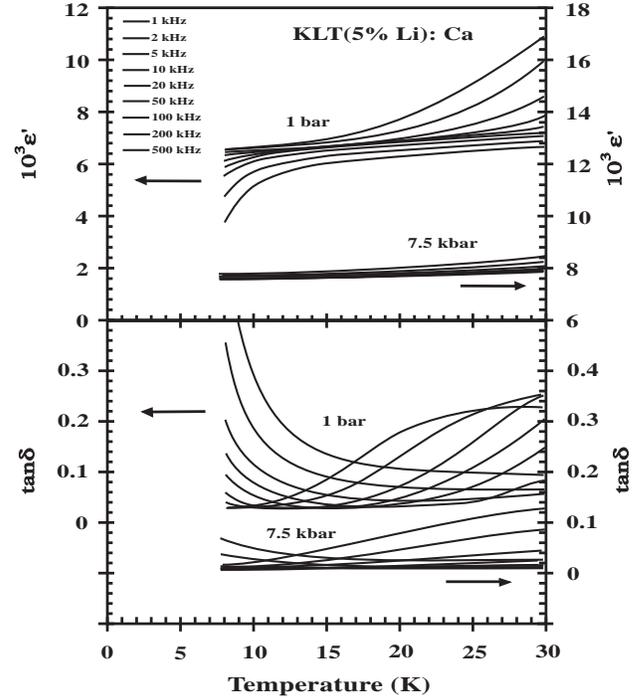}}
\caption{Low $T$ dielectric response at 1 bar and 7.5 kbar hints 
at the onset of a low $T$ ($< 8$~K) relaxation in KLT(5):Ca.}
\end{figure}

We speculate that the relaxation is, in fact, associated with the 
activation of the Ca-donated free carriers (electrons).  
Specifically, the carriers are frozen at the lowest temperatures and are 
activated at the onset of the relaxation.  To check this hypothesis, 
we have measured the dc resistivity ($\rho$) of our crystal (using In/Ga 
eutectic as contacts).  The results in Fig. 4 show that above the 
relaxation: $\rho < 10^5$ $\Omega$ cm, but that it increases sharply by 
over 3 orders of magnitude below 8~K, 
clear evidence of carrier freeze-out.  These results also 
support WempleÕs~\cite{Wemple_65} early finding for KTaO$_3$:Ca 
that carrier freeze-out occurs only at the lowest temperatures.  
The dielectric measurements were repeated at a pressure of 7.5~kbar.  
Pressure suppresses the frequency dispersion and shifts the 
relaxation to lower temperatures.  It also increases the 
amplitude of $\epsilon'(T)$ in the plateau region by 15~\% under 
7.5~kbar.  In early high pressure work on KT and other perovskites, 
Wemple {\it et al.}~\cite{Wemple_private} concluded that for these 
materials the electron mobility is related to $\epsilon'$ by the 
relationship $\mu \propto 1/\epsilon' + constant$.  
Since $\epsilon'$ decreases with pressure as a result of the 
increase in soft mode frequency, $\mu$ can be expected to increase 
leading to an increase in the conductivity.  It is this increase 
in sample conductivity that is responsible for the increase in 
the amplitude of $\epsilon'(T)$ in the plateau region with 
pressure.  Simply stated, the higher the conductivity of the 
sample, the greater is the conductivity mismatch at the barrier 
layer and the higher is the relaxation step in $\epsilon'$.~\cite{Grubbs_05}

\begin{figure}
\centerline{\epsfxsize=3.3in\epsfbox{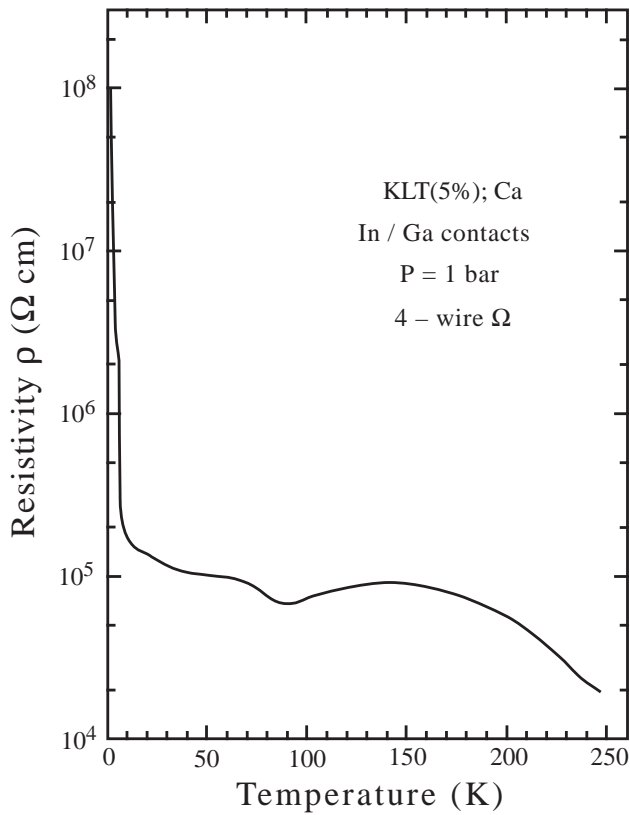}}
\caption{The temperature dependence of the resistivity, $\rho(T)$, 
of KLT(5):Ca showing carrier (electron) freeze-out below 10~K.  
It is the un-freezing of the carriers on heating that triggers 
the Maxwell-Wagner relaxation that results in the $\epsilon' = 6000$ 
plateau in Figs. 2 and 3.}
\end{figure}

A similarity between the results shown in Fig. 2 (a) for KLT(5):Ca and 
the undoped KLT(4.3) in the inset is represented by the shape of the $\epsilon'(T)$ 
response with its overlapping peaks, despite the fact that the 
absolute amplitudes of $\epsilon'(T)$ for the two materials are 
vastly different.  In fact, the response in Fig. 2 (a) is 
qualitatively like that of undoped KLT(5), but ``sitting on top of'' an 
$\epsilon' \sim 6000$~ baseline.  Remarkably, for the undoped 
crystal in the inset, $\epsilon'_{max}$ at the relaxational peak 
is $\sim 6 \times$ its value at low temperatures, and this ratio is about 
the same for the KLT(5):Ca crystal.  Pressure shifts the 
$\epsilon'(\omega,T)$ response to lower temperatures and sharpens 
it somewhat maintaining its relaxor character (Fig. 5).

\begin{figure}
\centerline{\epsfxsize=3.3in\epsfbox{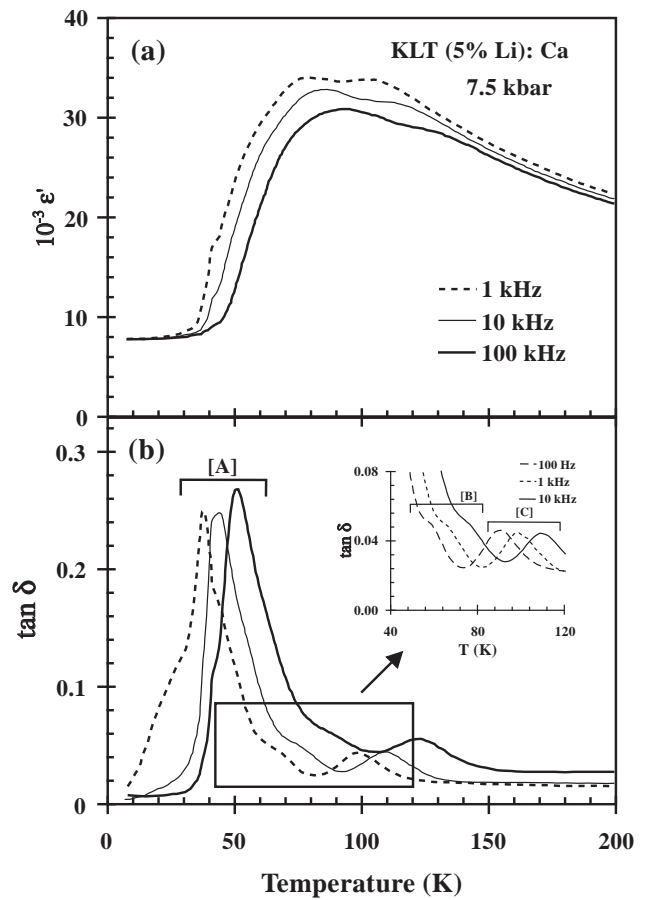}}
\caption{The dielectric $\epsilon'(T)$ and $\tan \delta$ responses 
of KLT(5):Ca at 7.5~kbar.  Pressure shifts the responses to lower $T$Õs 
and sharpens the relaxational features (compare to Fig. 2).}
\end{figure}

An additional feature in the $\epsilon'(T)$ response in Fig. 2 (a) 
is the structure at temperatures below $\epsilon'_{max}$.  This 
structure is revealed more distinctly in the $\tan \delta(\omega,T)$ 
response illustrated in Figs. 2 (b) and 5 (b).  Some of this response 
is probably due to some 
inhomogeneities or defects in the crystal, but there are three sets 
of relaxational peaks that we wish to draw attention to.  
These are labeled A, B and C in Figs. 2 (b) and 5 (b) 
and all are Debye-like, the relaxational frequency (or 
inverse relaxation time, $\tau$) obeying Arrhenius kinetics
\begin{equation}
\omega = \tau^{-1} = \omega_o \exp(-E/kT)
\end{equation}
as shown for A and C in Fig. 6.  The kinetic parameters are: $E = 79 (244)$~meV 
and $\omega_o = 4.2 \times 10^{12} (5 \times 10^{15})$~s$^{-1}$ for 
A (C), respectively.  These parameters are in close agreement with 
the above-cited values for the reorientation of the  isolated 
Li$^{+}$ and Li$^{+}$-Li$^{+}$ ion pair dipoles, respectively, in 
undoped KLT, and, thus, are unaffected by the light Ca doping.

The mechanism for the reorientation of the Li$^{+}$ dipole is by
$\pi/2$ $(90^{\circ})$ flips of the dipolar orientation among the 
six equivalent Li sites 
(Fig. 1 (b)).~\cite{Doussineau_93,Toulouse_98JKPS,Christen_91}  
At the 5~\% Li concentration in KLT(5):Ca, the Li$^{+}$ - Li$^{+}$ 
ion pairÕs signature is clearly present in both the $\epsilon'$ and 
$\tan \delta$ data.  The reorientation of this ion pair is believed 
to be associated with the reorientation of two nearest-neighbor Li$^{+}$ 
ions.~\cite{Doussineau_93,Toulouse_98JKPS,Wemple_private,Christen_91,Prosandeev_03}  
Doussineau {\it et al.}~\cite{Doussineau_93} have suggested that, 
in their two degenerate lowest energy configurations, the pair 
dipoles are collinear and aligned parallel to the line that joins 
their sites.  The relaxation mechanism involves a simultaneous 
reversal of both dipoles in $\pi$ (180$^{\circ}$) flips [Fig. 1 (c)].  
Support for this mechanism comes from the absence of this 
relaxation in ultrasonic measurements.~\cite{Doussineau_93,Christen_91}  
Because the postulated Li$^{+}$ - Li$^{+}$ pairs are centro-symmetric, 
a flip leaves the quadrupole moment invariant, and this relaxation 
mode is not coupled to acoustic waves.  Support has also come from 
recent first-principles calculations on KLT 
supercells.~\cite{Prosandeev_03}  These calculations indicate that 
the pairÕs 180$^{\circ}$ reorientation proceeds via metastable 
intermediate steps.  Additionally, the calculated hopping barriers 
for both the isolated Li$^{+}$ dipole and the pair are found to be 
in reasonable agreement with those deduced from the dielectric 
measurements.

Relaxation B appears as a peak in $\tan \delta$ at 100~Hz, but 
is a shoulder at other frequencies.  This shoulder becomes more 
prominent and more separated from the A and C relaxations at high 
pressure [Fig. 5 (b)].  An approximate separation of the B peaks, 
shows that this relaxation is also Debye-like with the same 
activation energy $E \sim 135$~meV.  We attribute this new B 
relaxation in KLT(5):Ca to the Ca$^{2+}$ dopant, since it is not 
observed in undoped KLT.  But in any case, it is a minor feature 
in the response of KLT(5):Ca.

\begin{figure}
\centerline{\epsfxsize=3.3in\epsfbox{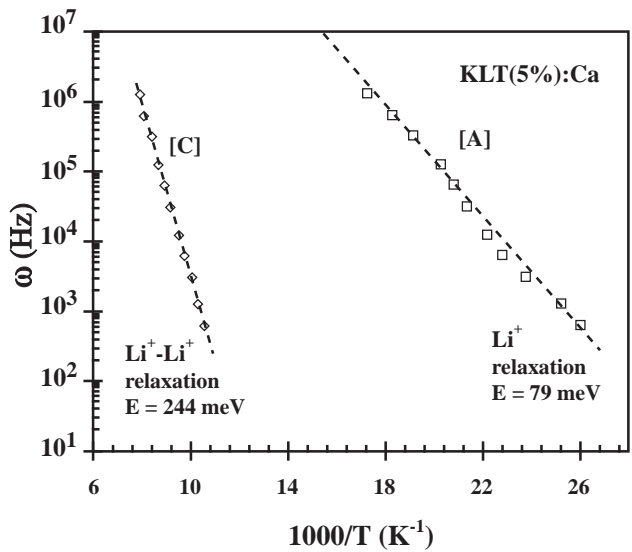}}
\caption{Arrhenius plots for the Li$^{+}$ [A] and Li$^{+}$-Li$^{+}$ 
pair [C] dipolar relaxations in KLT(5):Ca at 1~bar.}
\end{figure}

\subsection{Diffuse scattering and atomic shifts}

In the prototypical relaxor like Pb(Mg$_{1/3}$Nb$_{2/3}$)O$_3$ 
(PMN), it is well established that diffuse neutron 
scattering appears around the nuclear Bragg peaks below the 
so-called Burns temperature - below which the polar nano-regions
(PNR's) appear.~\cite{Naberezhnov_99}  Therefore, 
the diffuse scattering originates from the atomic displacements in 
PNR's, and its structure factor $F_{diff}({\rm \bf Q})$ can 
be written as
\begin{equation}
\label{f_diff} 
F_{diff}({\rm \bf Q}) = 
\sum_{\kappa} [{\rm\bf Q}\cdot \vec{\delta}_{\kappa}] 
b_{\kappa} e^{i{\rm\bf Q}\cdot{\rm\bf d}_{\kappa}} e^{-W_{\kappa}}, 
\end{equation}
where $\vec{\delta}_{\kappa}$, $b_{\kappa}$, ${\rm\bf d}_{\kappa}$, 
and $e^{-W_{\kappa}}$ are the atomic shift, neutron scattering 
length, coordination vector, and Debye-Waller factor 
(approximately 1) for $\kappa$ atoms, respectively.  
Note that the diffuse scattering intensity is proportional to $|F_{diff}|^2$.
For PMN, anomalous atomic shifts in the PNR's are reported; 
the total shift is composed of two components - one is 
a set of shifts that conserves the center-of-mass (CM) condition, 
and the other is a uniform phase shift.~\cite{Hirota_02}  
To verify if such anomalous shifts are also realized in the 
KLT system, we have carried out a study of the diffuse scattering from KLT(5):Ca.

\begin{figure}
\centerline{\epsfxsize=3.3in\epsfbox{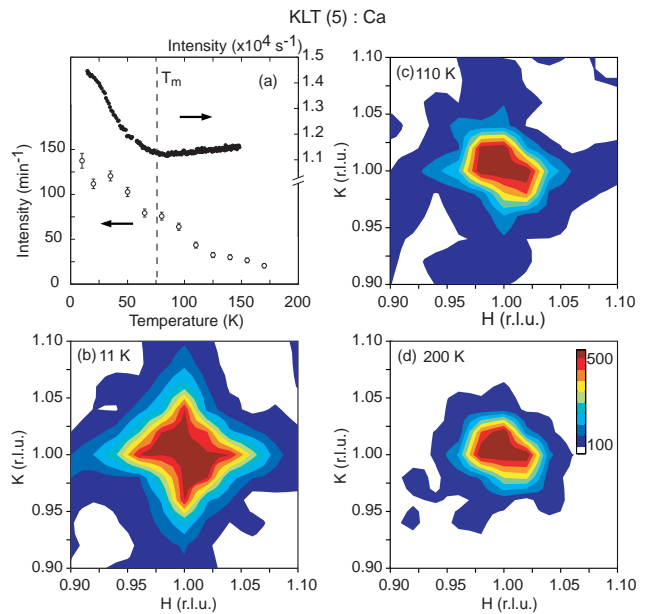}}
\caption{(Color online) (a) Temperature dependence of the $(2, 0, 0)$ 
nuclear Bragg peak (black circles) and the diffuse scattering 
intensity measured at $(1.05, 1, 0)$.  (b)-(d) Contour 
maps of the diffuse intensity around $(1, 1, 0)$ at 11~K, 
110~K, and 200~K.}
\end{figure}

The black circles in Fig. 7 (a) show the temperature dependence 
of the (200) nuclear Bragg peak.  Some relaxors that have an actual FE transition at 
$T_c$, such as KLT with higher Li concentration and Pb(Zn$_{1/3}$Nb$_{2/3}$)O$_3$ (PZN) 
have been reported to show an increase in the nuclear Bragg 
peaks below $T_c$ due to a release of extinction.  The 
present KLT(5):Ca crystal shows an increase of the (200) 
intensity below $75$~K.
Although our crystal does not show a sharp peak in the dielectric 
response at 75~K, this temperature is consistent with the first hump 
in the $\epsilon'(T)$ response in Fig. 2 (a) that corresponds to the 
slowing down of the fluctuations of the Li$^{+}$ dipoles.
Additionally, the increase of the nuclear Bragg intensity in the present 
crystal is only $\sim 50$~\% and it occurs gradually with decreasing 
temperature.  This is in contrast to the behavior of 
KLT crystals with higher Li content which have a well-defined 
FE transition in the dielectric measurements; for these latter crystals 
the nuclear peak 
intensity increases rapidly by more than an 
order of magnitude at $T_c$.~\cite{Toulouse_00,waki_unpub}  These facts 
suggest that the present crystal has a very small volume 
fraction of any ferroelectrically transformed area, and that a 
relaxor state is still dominant below 75~K, 
a conclusion supported by the dielectric results.
Hereafter, we define this temperature as $T_m$ (not $T_c$) since 
no bulk FE transition occurs in our crystal.

Temperature variations of the diffuse scattering around 
the (110) position are shown in Figs. 7(b) - 7(d).  
At $T=11$~K, diffuse scattering ridges along the $H$ and $K$ 
directions are clearly observed.  At $T=110$~K, which is 
$\sim 35$~K higher than $T_m$, the diffuse ridges are 
still visible, and then, the diffuse scattering finally becomes 
unobservable at 200~K.  The temperature dependence of this 
scattering intensity measured at $(1.05, 1, 0)$ is 
indicated in Fig. 7 (a) by open circles and shows gradual 
increase starting at $\sim 150$~K, 
suggesting that the Burns temperature of this system 
should be above this temperature.  The diffuse 
scattering does not show the critical scattering behavior 
around $T_m$, that is observed in relaxor materials 
with a well-defined $T_c$, such as PZN~\cite{Stock_PRB04} 
and KLT with higher Li content.~\cite{Toulouse_00}  
Again, this is supporting evidence that the present sample does not 
have a ferroelectrically well-ordered state.

\begin{figure}
\centerline{\epsfxsize=3.3in\epsfbox{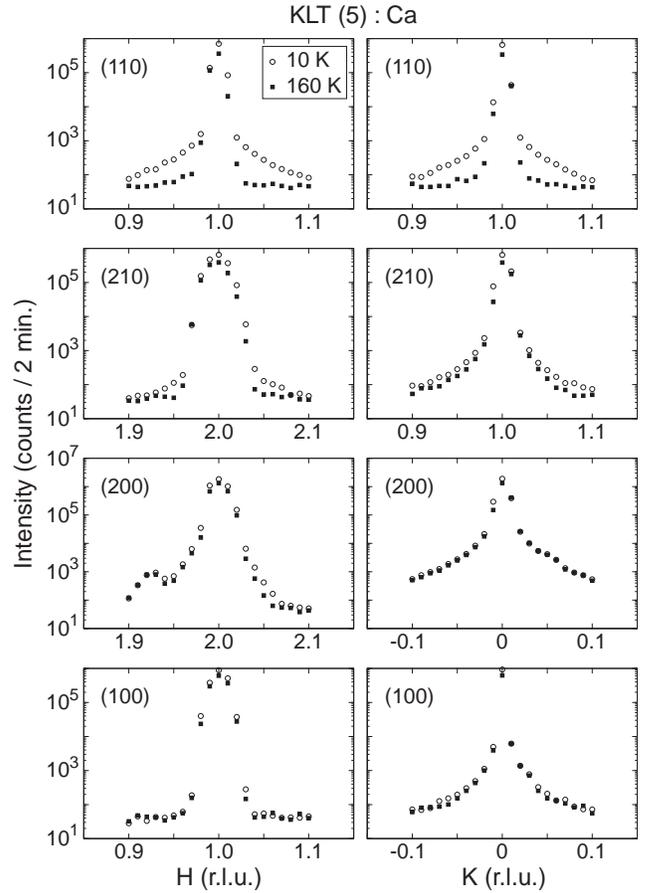}}
\caption{Comparison of diffuse scattering peaks at 10~K 
(open circles) and at 160~K (black circles) at various 
${\rm \bf Q}$ positions.  Left and right figures show 
scans along the $H$ and $K$ directions, respectively.}
\end{figure}

The observed diffuse scattering ridge along the [100] direction 
in the reciprocal space 
should correspond to a disk shape of PNR's which spreads in 
the $\{100\}$ plane in the real space, 
as discussed in Ref.~\onlinecite{Toulouse_00}.  
Similar {\it pancake} shaped PNR's in PMN and PZN-PT have been 
discussed in Ref.~\onlinecite{Xu_pancake}.  
From the width of the diffuse peaks, we estimate the dimensions 
of the disk PNR's as $52 (\pm 10)$~\AA~ in the \{100\} plane 
and $16 (\pm 3)$~\AA~ thickness in the [100] direction.

Diffuse scattering profiles at several zones are shown in Fig. 8.  
A clear diffuse component appears as wide intensity tails 
around (110), and a small diffuse intensity is 
observed around (210).  For (200), the longitudinal profile 
($H$-scan) is somewhat peculiar; the Bragg peak itself has 
already a wide tail even at 160~K.  (We note that a small 
hump at $H = 1.93$ is a powder line from the aluminum 
surrounding the sample.)  
The presence of this tail at this high temperature may be related to 
the dielectric results in Fig. 2 (a) that show relaxation 
associated with the Li$^{+}$-Li$^{+}$ dipolar pairs that 
extends to $\sim 200$~K.
We also find that there is an intensity 
difference between the results obtained at 160~K and 10~K at $H \sim 2.05$.  
This difference, however, is suppressed quickly at $H=2.07$, 
whereas the diffuse scattering decays gradually up to $q=0.1$.  
Therefore, we conclude that the difference in intensities at 
$H = 2.05$ is not associated with diffuse scattering 
but results from possible broadening of the Bragg peak 
in the longitudinal direction below $T_m$.  

Apparently, there is no clear diffuse signature around (100).  
Based on Eq.~\ref{f_diff}, structure factors for both 
(100) and (210) can be written as 
$F_{diff}({\rm \bf Q}) = [{\rm \bf Q} \cdot \hat{\delta}](A - B -O)$
where 
$A = 0.95 \delta_{\rm K} b_{\rm K} + 0.05 \delta_{\rm Li} b_{\rm Li}$, 
$B = \delta_{\rm Ta} b_{\rm Ta}$, 
$O = \delta_{\rm O} b_{\rm O}$, 
and $\hat{\delta}$ is a unit vector.
This means that the difference in the diffuse intensity 
between (100) and (210) arises from $|{\rm \bf Q}|^2$.  
Our observation demonstrates an existence of diffuse 
intensity at (210), thus, the diffuse intensity at 
(100) should be simply too small to observe rather than being zero.

In order to derive the atomic shifts by using Eq.~\ref{f_diff}, we employ 
the relation $F_{diff}(110) = 1.93 F_{diff}(210)$ and 
$F_{diff}(200) = 0$.  The former is based on a comparison 
of the diffuse intensities at $(1.05, 1, 0)$ and 
$(2.05, 1, 0)$.  
Then we assume $\hat{\delta}$ to be parallel to the Li off-center direction,
either $[100]$ or $[010]$.
Thus the local crystal symmetry inside the PNR is tetragonal.  
The PNR's with $\hat{\delta} = [100]$ and $[010]$ are also assumed 
to be distributed with same population.
These conditions give $A = -1.66 O$ and $B=-1.34 O$.  
Here we use the quantities $b_{\rm K}=0.35$, $b_{\rm Ta}=0.70$, 
and $b_{\rm O}=0.58$, and, for simplicity, neglect the 
$\delta_{\rm Li}$ contribution since the Li content (5~\%) 
is relatively low.  Then atomic shifts that are normalized to 
the shift of the oxygen atom ($\delta_{\rm O}$) are obtained as 
\begin{eqnarray*}
\delta_{\rm K} &=& -2.89,\\
\delta_{\rm Ta} &=& -1.11,\\
\delta_{\rm O} &=& 1.00.
\end{eqnarray*}

Remarkably, these values do not satisfy the 
CM condition, $\sum_{\kappa} \delta_{\kappa} M_{\kappa} = 0$, 
where $M$ is an atomic mass, $M_{\rm K} = 39$, 
$M_{\rm Ta} = 181$, and $M_{\rm O} = 16$.  
A similar breaking of the CM rule has been found in the 
atomic shifts of PMN.~\cite{Vakhrushev_95}  
Hirota {\it et al.}~\cite{Hirota_02} proposed a 
model of TO phonon condensation with a uniform 
phase shift.  In this model, the atomic shift can 
be expressed as 
$\delta_{\kappa} = \delta_{\kappa}^{cm} + \delta_{\kappa}^{us}$, 
where $\delta_{\kappa}^{cm}$ originates from the
condensation of the TO soft mode conserving the CM condition, 
whereas $\delta^{us}$ is a uniform shift of PNR's 
relative to the surrounding cubic matrix.  
Following the same procedure employed by Hirota {\it et al.}, 
we obtain
\begin{eqnarray*}
\delta_{\rm K}^{cm} &=& -1.90,\\
\delta_{\rm Ta}^{cm} &=& -0.12,\\
\delta_{\rm O}^{cm} &=& 1.99,\\
\delta^{us} &=& -0.99.
\end{eqnarray*}
%

\begin{figure}
\centerline{\epsfxsize=3.3in\epsfbox{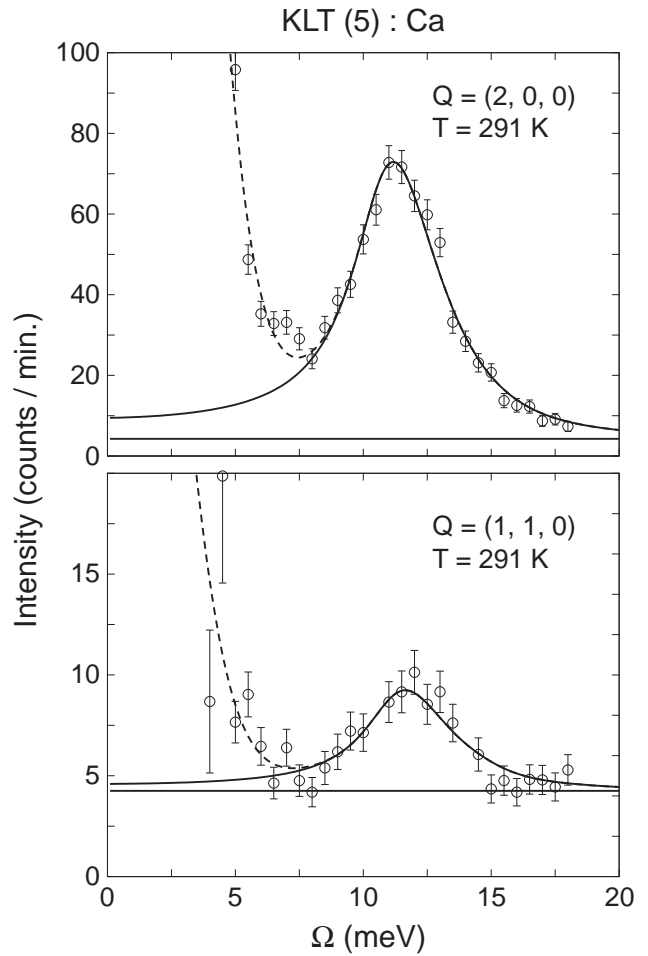}}
\caption{Zone center optic phonon profiles at $(2, 0, 0)$ 
(top) and $(1, 1, 0)$ (bottom).  Solid lines are fits to 
a resolution-convoluted Lorentzian function, while the dashed 
lines represent a Gaussian peak component at $\Omega = 0$.}
\end{figure}

It should be noted that, once the diffuse intensity ratios 
at different zones are given, the above quantities 
$\delta$, $\delta^{cm}$, and $\delta^{us}$ are uniquely 
derived only by a mathematical procedure.  
The validity of the model can be tested by checking to see if 
$\delta^{cm}$ can reproduce the intensity ratio of the 
zone center TO mode.  If $\delta_{\kappa}^{cm}$ 
originates from the condensation of the TO soft mode, 
it should be proportional to the phonon polarization 
vector $\vec{\xi}_{\kappa}$.  The inelastic structure factor 
$F_{inel}({\rm \bf Q})$ is expressed basically by the 
same formula as Eq.~\ref{f_diff} where $\vec{\delta}_{\kappa}$ 
should be replaced by $\vec{\xi}_{\kappa}$.  
From the obtained $\delta_{\kappa}^{cm} (\propto \xi_{\kappa}$), 
a ratio of the inelastic structure factors 
$|F_{inel}(200)|/|F_{inel}(110)|$ is calculated to 
be $\sim 2.9$.  The observed zone-center TO phonons at (200) and 
(110) are shown in Fig. 9 which were measured at 
room temperature.  The solid lines are fits to a Lorentzian 
phonon crosssection and a Gaussian peak at energy transfer $\Omega = 0$ 
convoluted with an instrumental resolution function.  (The actual 
formula is given in the next section.)  
Our fits give $|F_{inel}(200)|/|F_{inel}(110)| = 3.3 (\pm 0.3)$, 
which is in agreement with the value calculated 
from $\delta^{cm}$.

Another check of the consistency of the data may be achieved by determining the 
contributions of the Slater and Last modes, which are two 
dominant optic modes in perovskite materials.  In case 
of KLT, the Slater mode corresponds to an atomic motion 
whereby the oxygen ions and Ta ions move in opposite 
directions while the K ions do not move.  The Last mode 
corresponds to the counter displacements of the TaO$_{6}$ octahedra and K 
ions.  In order to satisfy the CM 
condition, the amplitudes $(\xi_{\rm K}, \xi_{\rm Ta}, \xi_{\rm O})$ 
for the Slater and Last modes should be $\vec{s}_1 = (0, -0.265, 1)$ 
and $\vec{s}_2 = (-5.872, 1, 1)$, respectively.  Designating the contributions 
of the Slater and Last modes to $S_1$ and $S_2$, 
respectively, we obtain 
$S_1 \vec{s}_1 + S_2 \vec{s}_2 = (\xi_{\rm K}, 
\xi_{\rm Ta}, \xi_{\rm O})$, giving $S_1 \sim 1.6$ 
and $S_2 \sim 0.3$.  Harada {\it et al.}~\cite{Harada_70} 
have reported an almost 100~\% contribution of the Slater 
mode in the non-Li-doped KT, which is qualitatively 
consistent with the present observation giving an 85~\% 
contribution by the Slater mode.  In the Li-doped samples, 
it is possible that the K-ions can move more since the doped 
Li ions with their smaller radius effectively provide more space.  
This is possibly why we have a somewhat larger contribution of 
the Last mode in KLT than in KT.
(In the same mannar, the small amount of Ca ions in our crystal, 
which shift to the off-center positions in combination with the 
oxygen vacancies, may also cause the same effect.)

The diffuse scattering in the relaxor KLT(5):Ca, therefore, exhibits 
analogous behavior to that of the prototypical relaxor PMN.  
Remarkably, the atomic shifts in KLT, that are characterized 
by the diffuse intensities, can be consistently 
explained by the uniform phase shift model introduced 
by Hirota {\it et al.}  Accordingly, it is likely that the uniform 
phase shift is a common feature in relaxor systems 
and is a key to understanding the fundamentals of the relaxor mechanism.

The origin of the uniform phase shift remains somewhat of an open question.  Hirota {\it et al.}~\cite{Hirota_02} have hypothesized that the phase shift in PMN occurs as the polarized regions slide on an electric field gradient which is due to the existence of chemical ordered and disordered domains.  In the ordered domain, Mg$^{5+}$ and Nb$^{2+}$ ions are ordered in 1:1 ratio, which breaks the charge neutrality in the domains, and thence the electric field gradient is induced in a comparable length scale to the PNRÕs.  Although there are no such the electric field gradient in KLT, as Li$^{+}$ and K$^{+}$ have the same valence, a somewhat analogous mechanism is likely operable.  In this case, the large off-center displacements of Li ions produce two dipolar entities that are sources of random electric fields that can trigger the phase shift. Thus, we envision that for both KLT and PMN electric fields trigger the phase shift which is then superimposed on the ionic displacements associated with the FE soft mode resulting in the total measured displacements within the PNRs.

\subsection{Lattice dynamics}

In relaxor materials, anomalies of the transverse optic (TO) and the 
transverse acoustic (TA) phonons are expected to accompany PNR formation.  
A feature that has recently attracted a significant amount of 
attention is the strong damping of the TO mode below 
the Burns temperature at small $q$, a feature 
observed in both PZN~\cite{Gehring_PRB01} and PMN~\cite{Gehring_PRL01}.  
This is interpreted as a result of PNR formation 
that disrupts the long wave-length, i.e. small-$q$, optic mode.  
Also, the line width of the TA mode has been found to 
broaden below the Burns 
temperature.~\cite{waki_02}  We have studied the TA and TO 
phonons in the present relaxor KLT(5):Ca crystal in order to understand 
the lattice dynamics and to learn if such phonon anomalies are universal 
features of relaxors.
It should be noted that the TO phonon in KLT originates mostly from the dynamics of the TaO$_{6}$ octahedra since the Slater mode is highly dominant as mentioned earlier.  On the other hand, the relaxational behavior of doped Li ions appears in the dielectric response as we see in Sec. IV A.

\begin{figure}
\centerline{\epsfxsize=3.3in\epsfbox{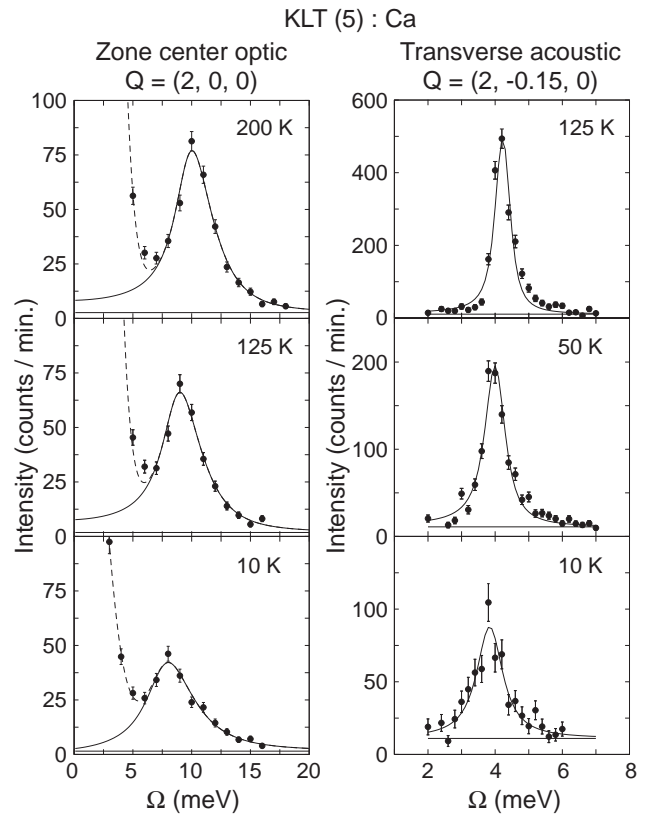}}
\caption{Zone center TO phonon measured at $(2, 0, 0)$ 
(left figures) and the TA phonon measured at 
$(2, -0.15, 0)$ (right figures).  Solid lines are fits to 
a resolution-convoluted Lorentzian function.}
\end{figure}

We have measured the phonons around the (200) Bragg peak since phonon 
structure factors at the other $Q$ positions that are accessible 
in the present condition of $E_f = 14.7$~meV are quite small.  
Figure 10 shows representative profiles of the TO phonons at the 
zone center (200) and of the TA phonons at $(2, -0.15, 0)$.   
It is seen that, with decreasing temperature, the phonon 
energy of the zone-center TO mode decreases and the line 
width apparently increases, but the phonon remains a well-defined 
excitation at all temperatures.  The TA phonon peak also broadens as 
the temperature decreases.

In order to extract more insight, all of the profiles have been analyzed by fitting to 
the resolution-convoluted phonon scattering function 
\begin{equation}
S({\rm \bf q}, \Omega) = [n(\Omega) + 1] \chi''({\rm \bf q}, \Omega),
\end{equation}
\begin{equation}
\chi''({\rm \bf q}, \Omega) = 
\frac{I}{(\Omega - \Omega_{ph}({\rm \bf q}))^2 + \Gamma (\Omega)} - 
\frac{I}{(\Omega + \Omega_{ph}({\rm \bf q}))^2 + \Gamma (\Omega)},
\end{equation}
where $\Omega$ is the neutron energy transfer $E_i - E_f$, 
$n(\Omega) = 1/(e^{\Omega / k_{B}T}-1)$ is the Bose factor, 
$\Omega_{ph}$ is the phonon energy, $\Gamma(\Omega)$ is the 
half-width-at-half-maximum (HWHM) of the phonon spectrum, 
and $I$ is the amplitude.  
To fit the TO spectra, the following dispersion relation was utilized;
\begin{equation}
\Omega_{ph}^2(q) = \Omega_0^2 + (Cq)^2,
\end{equation}
where $\Omega_0$ is a zone-center TO phonon energy, and 
$C = 40.3$~meV$^{2}$\AA$^2$ determined from the dispersion 
at room temperature.  For fitting of the TA profiles, 
the dispersion is taken to be linear in $q$.

\begin{figure}
\centerline{\epsfxsize=3.3in\epsfbox{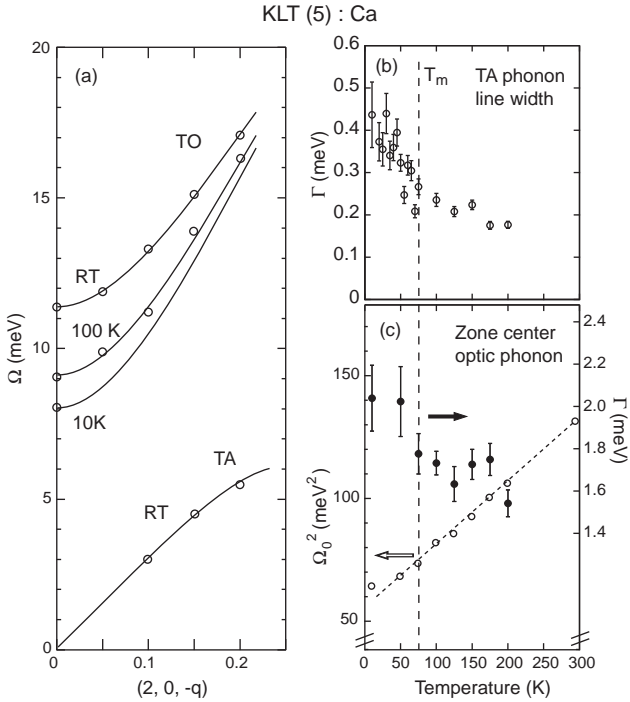}}
\caption{(a) Dispersions of the TO and 
TA phonons measured in the $(2, 0, 0)$ 
zone.  (b) Temperature dependence of the line width $\Gamma$ 
of the TA phonon measured at $(2, -0.15, 0)$.  
(c) Zone-center TO phonon energy squared $\Omega_0^2$ 
(open circles) and phonon line width $\Gamma$ (black circles) 
as a function of temperature.}
\end{figure}

The fits thus obtained are shown by solid lines in Fig. 10 
where the flat background is also adjusted.  Dashed lines 
represent elastic peaks at $\Omega = 0$ which are 
assumed to be Gaussians.
The fitting parameters are summarized in Fig. 11.  As 
shown in Fig. 11 (a), the TO mode softens gradually 
with decreasing temperature, but the dispersion is 
well-defined at all temperatures - in contrast to those 
of PMN and PZN.  The TO mode softening is 
clearly seen in Fig. 11 (c) where the zone-center TO 
phonon energy squared $\Omega_{0}^2$ plotted by 
open circles decreases linearly with temperature.  
This implies that the system still maintains an 
incipient lattice instability towards the 
ferroelectric state even with dopant Li ions present, 
similar to KT~\cite{Axe_70} and several KLT compositions.~\cite{Vogt_new}
Although the TO dispersion is well-defined at all 
temperatures, the TO phonon peak shows a small 
broadening at low temperatures.  The line width 
$\Gamma$ of the TO mode at the zone center, 
represented by the closed symbols in Fig. 11 (c), 
increases with decreasing temperature, 
characteristic soft mode behavior for ABO$_3$ perovskites.

The TA phonon dispersion at room temperature is 
shown in Fig. 11 (a).  We have observed a small 
softening by $\sim 1$~meV of the TA phonon energy at $q=0.15$~r.l.u. 
from room temperature to 10~K 
with a linear decrease in this energy.  
The TA phonon peaks are sharp and nearly 
resolution limited.  It is somewhat difficult 
to evaluate an intrinsic line width by 
deconvolution analysis which gives a large ambiguity 
for the resolution-limited peaks.  
Therefore, we have fit the TA profiles to a simple 
Lorentzian function to determine an effective line width.  
The line width, shown in 
Fig. 11 (b), increases with decreasing temperature 
even above $T_m$.  
This temperature-dependence resembles that of 
the diffuse intensity shown in Fig. 7 (a),
implying that the TA line broadening accompanies
the formation of PNR's.

There are interesting similarities and differences 
in the phonon behavior between the present KLT(5):Ca material 
and the prototypical relaxors PMN and PZN.  
In the case of PMN and PZN, the TO phonon softens 
following $\Omega_0^2 \propto T$ above the Burns 
temperature $T_d$.  
It then becomes strongly damped in the 
temperature range of $T_c ({\rm or}~ T_m) < T < T_d$, and finally recovers 
below $T_c$ ($T_m$) - and subsequently hardens with further decrease in 
$T$ - again following 
$\Omega_0^2 \propto T$.~\cite{Stock_PRB04,waki_02}
For the present KLT(5):Ca crystal, the TO phonon softens 
in the same manner, although it does not harden 
below $T_m$ because there is no macroscopically well-developed 
ferroelectric state.  
The softening at high 
temperatures for all these materials means that the relaxor state 
sets in ABO$_3$ lattices that have  
incipient lattice tendencies towards ferroelectric 
transitions. 
This is as expected because the high polarizability of these FE mode 
lattices results in large correlation lengths for dipolar interactions 
and favors the formation and growth of PNR's below the Burns 
temperature.~\cite{Grubbs_05}
No hardening of the TO phonon below 
$T_m$ in the present crystal means that the sample 
remains in the relaxor state at low temperatures.  
Consistent with this conclusion, the crystal shows neither a well-defined 
transition in the dielectric measurements nor a large 
extinction release of the nuclear Bragg peaks.

Another observed difference between KLT and PMN (PZN) is the absence of strong damping 
of the TO phonon in the present KLT(5):Ca crystal, 
which is frequently associated with the ``waterfall'' anomaly observed in PMN and PZN. 
We speculate
that this 
absence of strong damping 
occurs because the PNR's in the KLT(5):Ca crystal 
have small size and a small volume fraction.  Thus, these 
PNR's are not robust enough to 
dampen the TO phonon, but nevertheless can broaden the line width 
as shown in Fig. 11 (c). 
An alternative explanation for the waterfall effect has come from a recent neutron scattering study of 60\%-PbTiO$_{3}$ doped PMN by Stock {\it et al.}~\cite{Stock_05NWF} that revealed the  existence of the waterfall anomaly even though the sample exhibits long-range order,  suggesting that the waterfall is unrelated to PNR's.  They have argued that the waterfall results from random fields due to the chemical disorder (i.e., valence fluctuations).  It is important to note, however, that although mixed PbTiO$_3$-PMN samples exhibit FE order, they are highly, compositionally disordered and still exhibit considerable relaxor character.  If random fields are responsible for the waterfall effect, then it should be seen in KLT where the Li dipoles produce strong random fields.  Therefore our conclusion at this time is that the PNR explanation for the waterfall effect should not be dismissed.  However, further work is necessary to firm up the origin of this interesting effect.

As for the TA phonon behavior, in contrast to our KLT(5):Ca sample, 
Toulouse and 
Hennion~\cite{Toulouse_94} have reported that the TA 
phonon of KLT(3.5) splits into two modes below $T_c=52$~K corresponding 
to a tetragonal symmetry of the ferroelectric state.  
(They claim that their KLT(3.5) sample shows an FE transition 
at 52~K from the dielectric measurements.~\cite{Toulouse_94})
Our crystal shows only 
TA phonon broadening and no splitting because 
it does not have a ferroelectric transition.  
As shown in Fig. 11 (b), the TA broadening appears to 
start above $T_m$, and possibly at $T_d$ which 
should be around $200$~K as noted earlier based on the 
dielectric measurements and the diffuse scattering.  
Thus, the TA phonon is also affected by PNR 
formation.
In contrast to this behavior, PMN and PZN have been reported to 
show TA phonon broadening at temperatures in the range of 
$T_c < T < T_d$.~\cite{waki_02}  
Such behavior has been recently 
interpreted in the framework of a coupling of the TA mode 
to the diffuse component since the TA broadening is strong
at the {\rm\bf Q} positions where the diffuse scattering is 
strong.~\cite{Hlinka_03,Stock_cm04}  
More detailed 
measurements are necessary to determine if such coupling effects 
also exist in KLT as a manifestation of a common feature of relaxors.

Finally we discuss the TO phonon energy that depends on the Li 
concentration.  Comprehensive study of the TO soft mode using 
hyper-Raman spectroscopy for various Li concentrations have been 
carried out by Vogt.~\cite{Vogt_new}  It is clearly demonstrated 
that the TO mode frequency $\Omega_{0}$ increases with Li content 
$x$ following a linear relation 
\begin{equation}
\Omega_{0}^{2}(x, T) = \Omega_{0}^{2}(0, T) + A x.  
\label{Vogt_formula}
\end{equation}
In fact, the $\Omega_{0}$ value of our KLT(5):Ca crystal is higher 
than that of the pure KT measured by the neutron scattering by 
Axe {\it et al.}~\cite{Axe_70}  However, it is even higher than 
that of the $x=0.087$ sample measured by Raman spectroscopy by Vogt.  
To address this problem, we test the relation of 
Eq.~\ref{Vogt_formula} using the present KLT(5):Ca sample and 
another KLT crystal with $x=0.10$.

\begin{figure}
\centerline{\epsfxsize=2.8in\epsfbox{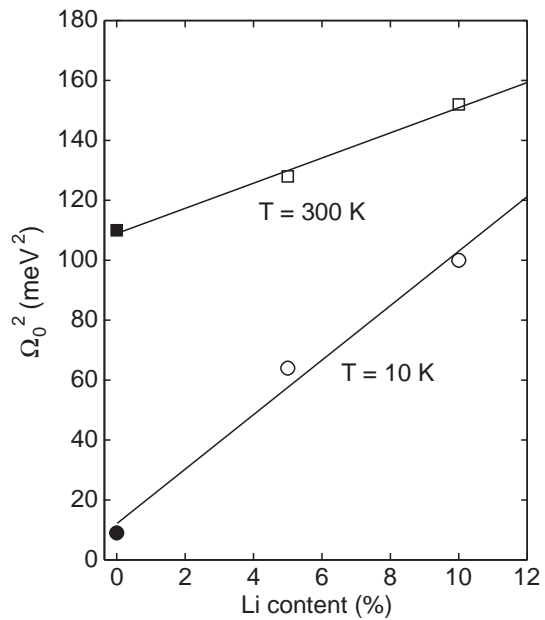}}
\caption{Li concentration dependence of $\Omega_{0}^{2}$ at $T=10$~K 
and $300$~K.  Data of pure KT (closed symbols) are adopted from 
Ref.~\onlinecite{Axe_70}.  The lines are results of fits to 
Eq.~\ref{Vogt_formula}.}
\end{figure}

The latter sample was grown in the same manner, but is free from 
Ca impurity.  It has a clear FE transition at $T_c=113$~K evidenced 
from dielectric measurements and neutron scattering experiments.  
(Details of this sample will be published elsewhere.~\cite{waki_unpub})  
The quantities of $T_m = 75$~K for KLT(5):Ca and $T_c = 113$~K for 
KLT(10) are consistent with {\it freezing temperatures}, $T_g$, 
summarized in Ref.~\onlinecite{Klink_83}, suggesting a validity of 
the estimation of the Li contents.

The zone-center TO phonon energy squared $\Omega_{0}^{2}(x, T)$ at 
$T=10$~K and $300$~K are shown in Fig. 12, together with those of 
pure KT referred from Ref.~\onlinecite{Axe_70}.  It is clearly 
demonstrated that the linear relation of Eq.~\ref{Vogt_formula} 
stands for the series of our samples.  This is further evidence 
of proper Li concentrations of our samples, and also, it implies 
that the small amount of Ca impurity has only a minor effect on 
the lattice dynamics properties.
We believe that the disagreement of $\Omega_{0}$ with those of 
Raman data by Vogt is due to a systematic difference of the Li 
content evaluation.

\section{Concluding remarks}

The origins of some aspects of the relaxor behavior in ABO$_3$ oxides remain as somewhat open questions.  For the prototypical relaxors PMN and PZN, however, it is clear that the chemical (valence) disorder of the Mg and Nb ions and the associated random fields play an important role by disrupting FE correlations leading to the relaxor state.  In the KLT system Li dipoles also produce random fields that are thought by some to be responsible for the relaxor behavior (e.g., see Ref.~\onlinecite{Kleemann_87}).  We note, however, that the situation in KLT is different from that in most ABO$_3$ relaxors in that the parent crystal KT has no FE order to disrupt.  In fact, at very low Li concentration in KT the Li dipoles exhibit essentially dipolar glass response with no correlations among dipoles.  Such correlations develop with increasing Li concentration leading to the evolution of a relaxor state

Combining the relaxational behavior and the neutron scattering results, we can draw the following picture for KLT(5).  At the Burns temperature, PNR's start to form triggered by off-center displacements of the Li ions.  The symmetry of these polar regions is tetragonal, and the observed ionic displacements consist of the displacements associated with the soft FE mode on which is superimposed a uniform phase shift.  

Below the Burns temperature, the dynamics of Li atoms appears as the relaxational behavior in the dielectric response. As temperature decreases dipolar fluctuations are reduced and stronger correlations develop among the PNRs that grow in volume, as evidenced by the growth of diffuse scattering intensity. Slowing down of dipolar fluctuations sets in at a temperature corresponding to the peak temperature in the dielectric constant and continues on further cooling.  Although the present sample shows no bulk FE transition, the increase of the Bragg intensity below $\sim 75$~K, where the Li-Li pair becomes well static ($1000/T \sim 13$ in Fig. 6) is indicative of larger PNRs  (which are nano FE domains) or even the presence of larger FE domains in a sea of relaxor phase.

In summary, 
we have measured the dielectric response and lattice dynamics 
in the relaxor KLT(5):Ca where a small amount of Ca was incorporated 
into the KLT crystal.  
Both types of results affirm that the present crystal remains in the 
relaxor state and does not develop a ferroelectric state at 
low temperatures.  Dielectric constants that are sensitive 
to impurities exhibit strong enhancement attributed to free 
electrons from the Ca dopant, however 
the relaxational features including Arrhenius parameters of 
Li$^{+}$ ion and Li$^{+}$-Li$^{+}$ pair dipolar relaxations 
are not affected by the Ca.  

Neutron scattering results for KLT(5):Ca compared with those 
of the relaxor PMN highlight common features in relaxors.  
Lattice displacements in PNR's derived from the diffuse 
intensities suggest that the model of the normal soft-mode 
condensation with an additional uniform phase shift within 
the PNR's is also relevant in KLT.  
This means that we cannot rule out a contribution of the 
zone-center TO soft mode to the formation of PNR's.
However, the TO soft mode remains at relatively high energy compared to 
that of PMN, and it also remains well-defined in contrast to the heavily-damped 
nature characteristic of PMN.
We believe that this difference reflects the fact that KLT(5):Ca 
is a weak relaxor, essentially a dipolar glass, whereas PMN is 
the prototypical strong relaxor. 
A more comprehensive study will be required in order to 
extract the precise nature of the specific coupling 
between the TO phonon and the relaxor mechanism in this system.

\begin{acknowledgments}

The authors thank B. Burton, T. Egami, R. Erwin, P. M. Gehring, 
K. Hirota, K. Kakurai, and N. Metoki 
for invaluable discussions.  
This work was partially supported by the US-Japan Cooperative 
Research Program on Neutron Scattering. 
Work at Sandia National Laboratory was supported by 
the Division of Material Sciences and Engineering, Office of Basic 
Energy Sciences, U.S. Department of Energy under Contract 
DE-AC04-94AL85000.  
Research sponsored in part by the Division of Material Sciences 
and Engineering, Office of Basic 
Energy Sciences, U.S. Department of Energy under Contract 
DE-AC05-00OR22725 with Oak Ridge National Laboratory, managed and
operated by UT-Battelle, LLC.
Finally, work at Brookhaven National Laboratory is supported by 
financial support from the U. S. Department of Energy under 
Contract No. DE-AC02-98CH10886.

\end{acknowledgments}



\begin{references}



\bibitem{Wemple_65} S. H. Wemple, Phys. Rev. {\bf 137}, A1575 (1964).

\bibitem{Axe_70} J. D. Axe, J. Harada, and G. Shirane,
Phys. Rev. B {\bf 1}, 1227 (1970).

\bibitem{Raman} P. Calvi, P. Camagni, E. Giulotto, and L. Rollandi, 
Phys. Rev. B {\bf 53}, 5240 (1996).

\bibitem{Vogt_new} H. Vogt, J. Phys:Cond. Matter {\bf 7}, 5913 (1995); 
Ferroelectrics {\bf 184}, 31 (1996); {\bf 202}, 157 (1997).

\bibitem{Prater_81} R. L. Prater, L.L. Chase and L. A. Boatner, 
Phys. Rev. B {\bf 23}, 5904 (1981).

\bibitem{Toulouse_00} G. Yong, J. Toulouse, R. Erwin, S. M. Shapiro,
and B. Hennion, Phys. Rev. B {\bf 62}, 14736 (2000).

\bibitem{Naberezhnov_99} A. Naberezhnov, S. Vakhrushev, B. Doner, D.
  Strauch, and H.  Moudden Eur. Phys. J. B {\bf 11}, 13 (1999)

\bibitem{Vakhrushev_95} S. B. Vakhrushev, A. A. Naberezhnov, 
N. M. Okuneva, and B. N. Savenko, Fiz. Tverd. Tela (St. Petersburg) 
{\bf 37}, 3621 (1995) [Phys. Solid State {\bf 37} 1993 (1995)].

\bibitem{Hirota_02} K. Hirota, Z.-G. Ye, S. Wakimoto, P. M. Gehring, 
and G. Shirane, Phys. Rev. B {\bf 65}, 104105 (2002).

\bibitem{Lynn_note} We have performed inductively coupled plazma (ICP) analysis on the starting powder materials.  The following metals were sought: Ca, Bi, Cr, Fe, Mn, Sr, V. Sb, B, Co, Pb, Mo, Si, Te, Zn, As, Cd, Cu, Ni, Ag, Sn, Zr, Ba, In, Mg, P, Na, and Ti.  The ICP analytical results did reveal the presence of $< 15$~ppm of only Ca in the potassium carbonate.

\bibitem{Klink_84} J. J. van der Klink and F. Borsa, 
Phys. Rev. B {\bf 30}, 52 (1984).

\bibitem{Samara_SSP} G. A. Samara in Solid State Physics, 
Vol.56 edited by H. Ehrenreich and F. Spaepen, Academic Press, 
New York (2001) p. 239 and references therein.

\bibitem{Hochli_90} U. T. Hochli, K. Knorr and A. Loidl, 
Adv. Phys. {\bf 39}, 405 (1990).

\bibitem{Yacoby_74} Y. Yacoby and S. Just, 
Solid State Commun. {\bf 12}, 715 (1974).

\bibitem{Leung_01} K. Leung, Phys. Rev. B {\bf 65}, 012102 (2001). 

\bibitem{Senhouse_66} L.S. Senhouse, Jr., M.V. DePaolis, Jr. and 
T.L. Loomis, Appl. Phys. Lett. {\bf 8}, 173 (1966).

\bibitem{Prosandeev_01} S. A. Prosandeev, V.A. Trepakov, 
M.E. Savinov, J. Jastradik and S.E. Kapphan, 
J. Phys.: Condens. Matter {\bf 13}, 9749 (2001).

\bibitem{Doussineau_93} P. Doussineau, Y. Farssi, C. Frenos, 
A. Levelut, K. McEnaneu, J. Toulouse and S. Ziolkiewicz, 
Europhys. Lett. {\bf 24}, 415 (1993). 

\bibitem{Toulouse_98JKPS} J. Toulouse and R. Pattnaik, 
J. Korean Phys. Soc. {\bf 32}, S942 (1998).

\bibitem{Grubbs_05} R. K. Grubbs, E. L. Venturini, P. G. Clem, 
J. J. Richardson, B.A. Tuttle and G. A. Samara, 
Phys. Rev. B {\bf 72}, 104111 (2005).  
See also J. Volger in {\it Progress in Semiconductors}, edited by 
A. F. Gibson (John Wiley and Sons, New York 1960), vol. 4, p. 207.

\bibitem{Wemple_private} S. H. Wemple, M. DiDomenico, Jr. 
and A. Jayaraman, Phys. Rev. {\bf 180}, 547 (1969).

\bibitem{Christen_91} H. M. Christen, U.T. Hochli, A. Chatelain 
and S. Ziolkiewicz, J. Phys.: Condens. Matter {\bf 3}, 8387 (1991).

\bibitem{Prosandeev_03} S.A. Prosandeev, E. Cockayne and B.P. Burton, 
Phys. Rev. B {\bf 68}, 014120 (2003).


\bibitem{waki_unpub} S. Wakimoto, G. A. Samara, and L. A. Boatner, unpublished.

\bibitem{Stock_PRB04}C. Stock, R.J. Birgeneau, S. Wakimoto, J.S. Gardner, 
W. Chen, Z.-G. Ye, and G. Shirane, Phys. Rev. B {\bf 69}, 094104 (2004).

\bibitem{Xu_pancake} G. Xu, Z. Zhong, H. Hiraka, and G. Shirane,
Phys. Rev. B {\bf 70}, 174109 (2004).

\bibitem{Harada_70} J. Harada, J. D. Axe, and G. Shirane, Acta Cryst. 
{\bf A 26}, 608 (1970).

\bibitem{Gehring_PRB01} P. M. Gehring, S.-E. Park, and G. Shirane,
Phys. Rev. B {\bf 63}, 224109 (2001).

\bibitem{Gehring_PRL01} P. M. Gehring, S. Wakimoto, Z.-G. Ye, and G. Shirane,
Phys. Rev. Lett. {\bf 87}, 277601 (2001).

\bibitem{waki_02} S. Wakimoto, C. Stock, R. J. Birgeneau, Z.-G. Ye, 
W. Chen, W. J. L. Buyers, P. M. Gehring, and G. Shirane,
Phys. Rev. B {\bf 65}, 172105 (2002).

\bibitem{Toulouse_94} J. Toulouse and B. Hennion,
Phys. Rev. B {\bf 49}, 1503 (1994).

\bibitem{Hlinka_03} J. Hlinka, S. Kamba, J. Petzelt, J. Kulda, 
C. A. Randall, and S. J. Zhang,
J. Phys.: Condes. Matter {\bf 15}, 4249 (2005).

\bibitem{Stock_cm04} C. Stock, H. Luo, D. Viehland, J. F. Li, 
I. Swainson, R. J. Birgeneau, and G. Shirane,
J. Phys. Soc. Jpn. {\bf 74}, 3002 (2005).


\bibitem{Klink_83} J. J. van der Klink, D. Rytz, F. Borsa, and 
U. T. H{\"o}chli, Phys. Rev. B {\bf 27}, 89 (1983).

\bibitem{Stock_05NWF} C. Stock, D. Ellis, I. P. Swainson, G. Xu, H. Hiraka, Z. Zhong, H. Luo, X. Zhao, D. Viehland, R. J. Birgeneau, and G. Shirane, Phys. Rev. B {\bf 73}, 064107 (2006).

\bibitem{Kleemann_87} W. Kleemann S. K\"utz, and D. Rytz, Europhys. Lett. {\bf 4}, 239 (1987).


\end{references}
\end{document}